\documentclass[conference]{IEEEtran}
\IEEEoverridecommandlockouts
\usepackage{cite}
\usepackage{amsmath,amssymb,amsfonts}
\usepackage{algorithmic}
\usepackage{graphicx}
\usepackage{textcomp}
\usepackage{xcolor}

\usepackage[caption=false]{subfig}

\begin{document}

\title{5G meets Construction Machines: \\ Towards a Smart working Site}
\makeatletter
\newcommand{\linebreakand}{%
  \end{@IEEEauthorhalign}
  \hfill\mbox{}\par
  \mbox{}\hfill\begin{@IEEEauthorhalign}
}

\makeatother

\author{
  \IEEEauthorblockN{1\textsuperscript{st} Yusheng Xiang}
  \IEEEauthorblockA{\textit{Institute of Vehicle System Technology} \\
    \textit{Karlsruhe Institute of Technology}\\
    Karlsruhe, Germany \\
    yusheng.xiang@partner.kit.edu}
  \and
  \IEEEauthorblockN{2\textsuperscript{nd} Bing Xu }
  \IEEEauthorblockA{\textit{Institute of Vehicle System Technology} \\
    \textit{Karlsruhe Institute of Technology}\\
    Karlsruhe, Germany \\
    uxuko@kit.edu}
  \and
  \IEEEauthorblockN{3\textsuperscript{rd} Tianqing Su}
  \IEEEauthorblockA{\textit{Institue of Communication Technology} \\
    \textit{Technical University of Braunschweig}\\
    Brunswick, Germany \\
    t.su@tubs.de}
  \linebreakand 

  \IEEEauthorblockN{4\textsuperscript{th} Christine Brach}
      \IEEEauthorblockA{\textit{Division of Mobile Hydraulics} \\
    \textit{Robert Bosch GmbH}\\
    Elchingen, Germany \\
    christine.brach@boschrexroth.de}
  \and
  \IEEEauthorblockN{5\textsuperscript{th} Samuel S.  Mao}
  \IEEEauthorblockA{\textit{Department of Mechanical Engineering} \\
    \textit{University of California, Berkeley}\\
    Berkeley, USA \\
    ssmao@berkeley.edu}
  \and
  \IEEEauthorblockN{6\textsuperscript{th} Marcus Geimer}
  \IEEEauthorblockA{\textit{Institute of Vehicle System Technology} \\
    \textit{Karlsruhe Institute of Technology}\\
    Karlsruhe, Germany \\
    marcus.geimer@kit.edu}
}
\maketitle

\begin{abstract}
The fleet management of mobile working machines with the help of connectivity can increase safety and productivity. Although in our previous study, we proposed a solution to use IEEE 802.11p to achieve the fleet management of construction machines, the shortcoming of WIFI may limit the usage of this technology in some cases. Alternatively, the fifth-generation mobile networks (5G) have shown great potential to solve the problems. Thus, as the world's first academic paper investigating 5G and construction machines' cooperation, we demonstrated the scenarios where 5G can have a significant effect on the construction machines industry. Also, based on the simulation we made in $ns-3$, we compared the performance of 4G and 5G for the most relevant construction machines scenarios. Last but not least, we showed the feasibility of remote-control and self-working construction machines with the help of 5G.  

\end{abstract}


\section{Introduction}
The fleet management of mobile machines is the principal research direction of the internet of things in the construction machines industry. Besides using the ad-hoc network as the first version for mobile machines \cite{Xiang.2020b}, 5G attracts huge attention to be expected to achieve even higher-quality communication. 
As mentioned in our previous study, WIFI technology can accomplish realtime communication among mobile machines so that they will work denser and faster. As a consequence, we can increase productivity and therefore reduce the duration of the construction projects. This is meaningful for the cases of repairing projects on the highway, mining projects, and transportation in harbors. Since mobile machines are usually working surrounded by dust and Lidars are quite sensitive to this case, cameras are a more robust and promising approach towards self-working machines or remote control of mobile construction machines. As we know, as the videos' resolution increases, both image recognition algorithms, and humans can acquire information easier and more accurate. However, the capacity, especially the uplink capacity of WIFI technology, limits the introduction of wireless high-definition (HD) video transmission for construction machines.  As we did not find comprehensive research indicating how can 5G change the mobile construction machines industry, we first analyze the potential use cases for the implementation of 5G for the construction machines industry in this paper. Followed by illustrating the benefits by utilizing 5G with our simulation results by means of $ns-3$ \cite{Henderson.2008}. Last but not least, we show the blueprint of future smart working sites based on the simulation results. Fig. \ref{fig: remote control} and Fig. \ref{fig: self-working} demonstrate the potential use cases of 5G in the field of mobile construction machines. 

\begin{figure}[htbp]
\centerline{\includegraphics[width=3.0in]{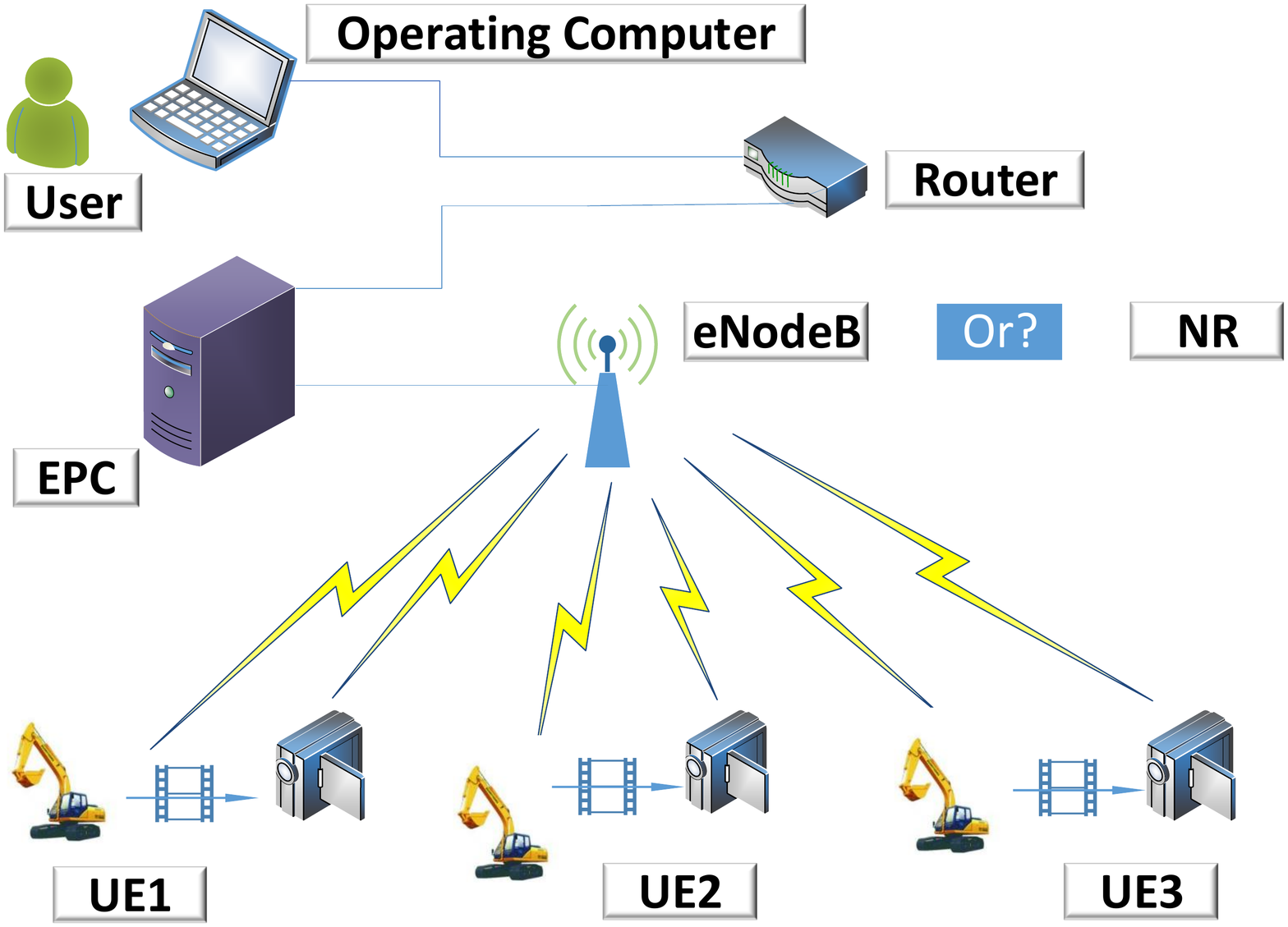}}
\caption{Remote control with live streaming: here cameras will be installed on the mobile machines while the driver sits in a comfortable room to operate the machines remotely. Thanks to 5G, HD video streaming can be sent with low delay and high reliability.}
\label{fig: remote control}
\end{figure}

\begin{figure*}[htbp]
\centerline{\includegraphics[width=7in]{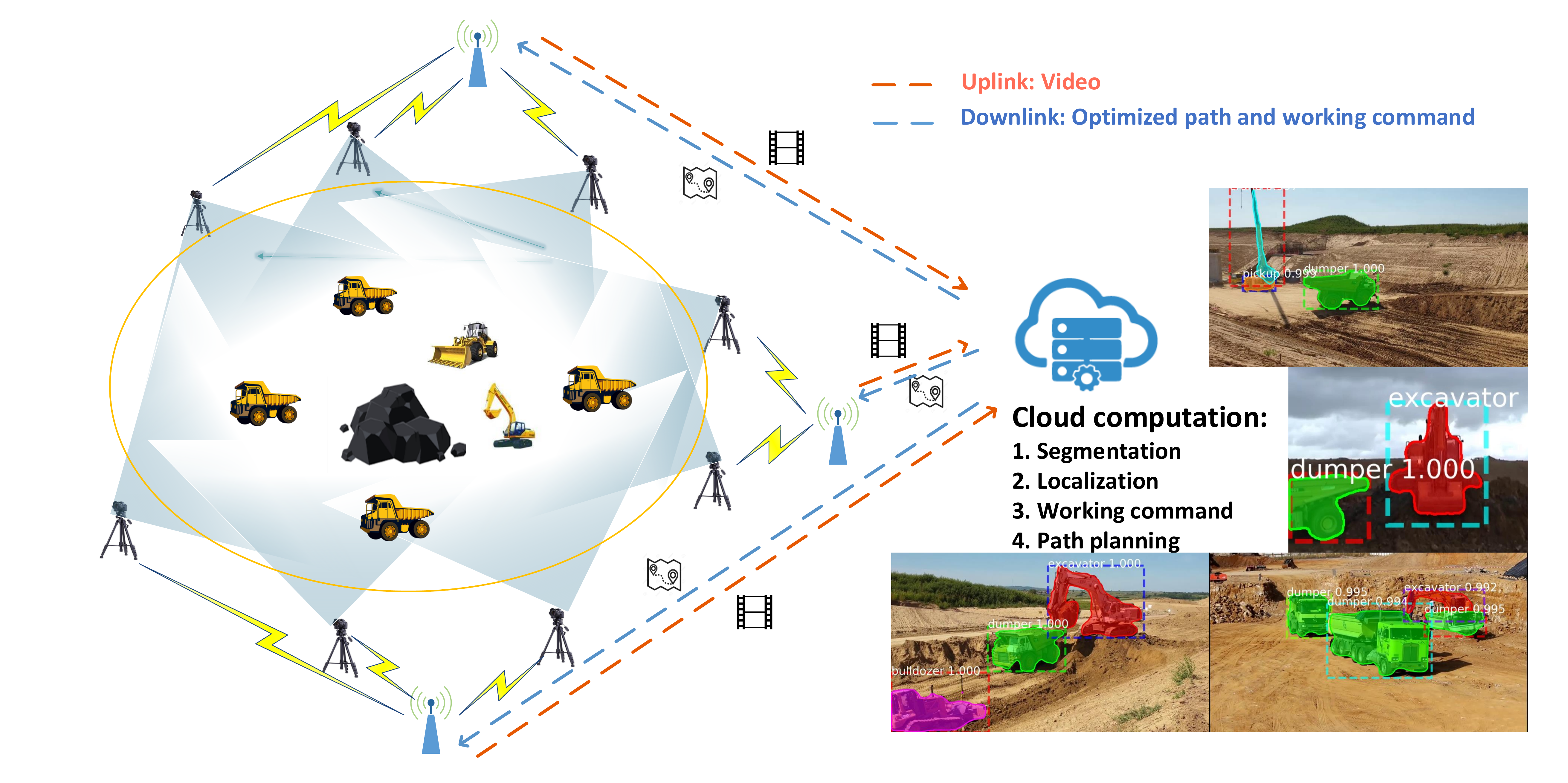}}
\caption{Self-working mobile machines: here, cameras will be fixed on the ground instead of being installed on the machines to avoid the obstruction of vision. The stream will be uploaded to the center commander and be processed on the cloud. Based on the stream from more than two cameras, the depth information and motion of machines can be acquired. Afterward, the command signal will be sent directly to the machines. The research about instance segmentation of construction machines can be found in \cite{Xiang.2020e}.}
\label{fig: self-working}
\end{figure*}

\section{current 5G technologies}
2020 is considered the first year of the 5G era in the wireless community since 5G is commercially employed in this year. To date, 5G is still a fast-developing research subject; thus, opposite views exist simultaneously. To avoid exaggerate the 5G technology, we only take the parameters and data that more than at least half of the community agree with into account.  

To overcome the shortcoming of 4G \cite{Ghosh.2010}, the basic requirements for the 5G are drawn by \cite{Chen.2014, Gupta.2015,Lauridsen.2017,Tahir.2019, Vinel.2017}: higher transmission rate, shorter latency, higher reliability, and more User Equipment (UE) connection. Correspondingly, the big 3 concepts: enhanced Mobile Broadband (eMBB), Ultra Reliable Low Latency Communications (URLLC), massive Machine Type Communications (mMTC) \cite{Shafi.2017}, were proposed. 
According to the 3rd Generation Partnership Project (3GPP) 38.101 agreement \cite{.0829202022:00:45}, 5G NR mainly uses two frequency bands: FR1 frequency band and FR2 frequency band. The frequency range of the FR1 band is 450MHz-6GHz, which is also called the sub 6GHz sub frequency band; the frequency range of the FR2 band is 24.25GHz-52.6GHz, usually called millimeter wave (mmWave) band. Currently, the most influential providers in the field of 5G are Huawei for sub 6Ghz band and Qualcomm for the mmWave band, separately. Other competitors mentioned quite often are Samsung, Ericsson, Datang, Nokia, Telecom, Intel, and ZTE. As we know, the higher the frequency, the closer the characteristic is to the light. That is, the propagation of the signal will be more similar to the light, which only goes straightforward so that the obstacles can easily block it. Also, the energy loss increases dramatically as the propagation distance increases, and proportionally to the square of the frequency. Consequently, the coverage problem, which restricts the promotion of the high-frequency spectrum 5G, occurs due to the nature of the mmWave. For this reason, most countries, such as China, Japan, and Korea, give priority to the sub 6Ghz band since the coverage is much larger, and thus more people can benefit from 5G technology. Compared to 4G, which only has 20MHz channel bandwidth, 5G is allocated about 100MHz in the sub 6Ghz area. Moreover, thanks to the novel Multiple-Input Multiple-Output (MIMO) technology, more antennas are used simultaneously to achieve a much higher transmission rate than the previous 4G technology. Compared to the 4G handsets, which only have 2*2 or 4*4 antennas, 5G base stations and UEs have antenna array to increase the spectrum utilization \cite{Deng.2017, Li.2018}. However, since such 5G UEs also use the sub 6Ghz band, there is principally not greatly different than 4G, and thus some serious problems are inevitable.  First of all, because the sub 6Ghz area is also used by 2G, 3G, 4G, and thus already very crowed, a further increase of the bandwidth is almost impossible. Although some communication operators give 5G more channel bandwidth, which was belongs to 2G and 3G to increase the bandwidth of 5G further, the bandwidth is surely not enough for the future potential requirements. In addition, the configuration of the antenna depends on the signal frequency. At sub 6Ghz, the wavelength is more than 1cm, so that the number of the antenna in the UE, in this case, is also limited. Therefore, soon after sub 6Ghz was promoted, how to use the higher FR2 frequency regions, i.e., higher than 28Ghz, has become a hot topic. Compared to the sub 6Ghz region, it is quite easy to have 1Ghz channel bandwidth in the FR2 region so that the transmission ratio is expected to be much higher. In the mmWave frequency band, taking the 28GHz frequency band as an example, the available spectrum bandwidth has reached 1GHz, while the available signal bandwidth of each channel in the 60GHz frequency band is 2GHz \cite{.0829202022:00:45}. In the case of constant spectrum utilization, if the mmWave frequency band is selected, the data transmission rate can be doubled by directly doubling the bandwidth. Since 3GPP has decided to continue to use orthogonal frequency division multiplexing (OFDM) technology for 5G NR \cite{.0829202022:00:45}, mmWave technology has become the biggest novel idea of 5G. Although mmWave is already used by satellite, they were considered as infeasible for the daily life scenarios. Until recently, the novel technology unlocks the high-frequency spectrum. Concretely, thanks to antennas array, which constitutes a large number of antennas and the beamforming technology \cite{Han.2015}, the energy can be concentrated in small regions. Moreover, because the antennas for mmWave can be designed much smaller than the microwave antennas, the antennas in the mmWave antenna array are much denser and achieve a larger number for the same geometrical apparatus. Along with a certain number of small cell base stations, mmWave comes to the forefront of commercial applications. The introduction of other important 5G technologies, such as new numerology, LDPC/Polar codes, etc., can let OFDM technology better extend to the mmWave band. To adapt to the large bandwidth characteristics of mmWave, 5G defines multiple sub-carrier intervals, of which the larger sub-carrier intervals are specifically designed for mmWave, whereas the lower is for the compatibility of previous system deriving from the 4G era. One of the main goals of 5G is to support URLLC services with stringent requirements for reliability and delay. LTE achieves a user plane two-way wireless delay of less than 10ms, and the design goal of 5G is to reduce this delay by at least 5 times, that is, less than 2ms. According to the 3GPP TS 38.211 protocol \cite{.0829202022:00:45}, the 5G NR physical layer provides multiple sub-carrier spacing configurations \cite{Campos.2017}. By increasing the sub-carrier spacing, the duration of OFDM symbols is reduced, thereby reducing the duration of a single time slot and reducing delay. The 3GPP protocol claims that the sub-carrier spacing is inversely proportional to the OFDM symbol duration, which is an inherent attribute of OFDM. For the current network communication technology, the key capability indicators of the 5G system have been greatly improved. The information transmission delay of the 5G network can reach milliseconds, which meets the stringent requirements of the network and guarantees the safety of controlled UE. The peak rate of 5G can reach 10-20 Gbit/s, and the number of connections can reach 1 million/$km^2$ \cite{Mohyeldin.2016}. 
Apparently, although the technology can overcome the difficulties of implementing the mmWave, the base stations for mmWave are energy-consuming equipment. Thus, Heterogeneous Network (HetNet) is also essential in the 5G era, i.e., most scientists in the wireless community believe that both sub- and above 6Ghz networks will coexist in a long time. The same as LTE, 5G also has device to device network to solve the problem when UEs are outside of the coverage of base stations \cite{Ansari.2018}.

\section{Where can construction sites be benefited from 5G?}

According to GSMA's outlook, mmWave can roughly make economic benefits 212 billion dollars only in the Asia Pacific region in 2034. Among them, 3\% to 9\% of the amount will come from the agriculture and mining industry. 

Although 5G shows excellent progress compared to 4G and WIFI, for end customers to accepted a new technology, a sudden colossal improvement is always necessary. Currently, most people believe that IoT technologies will endow the mobile construction machines industry with the ability, such as predictive maintenance, data analytics, and visualization and notification. However, we find that they are actually nice-to-have technologies. Since 5G may need a lot of micro base stations, and they are also energy-consuming \cite{Roh.2014}, the value created by predictive maintenance is quite difficult to compensate for the additional cost of 5G. In many cases, preparing some backup vehicles can be a more effective and money-saving solution. Moreover, the shortcoming of mmWave will be amplified by the harsh environment on the working site, such as the blockage of dust and giant machines. Thus, we believe more realistic scenarios are remote control and self-working mobile machines since 5G achieves something we cannot do well before. In some dangerous traditional industries, such as remote maintenance of underground pipelines, remote rescue of landslides, underground mine excavation, etc., these industries' operating environment is hazardous and harmful to the human body. Although remote control is achieved with a wired network for nowadays projects, the flexibility is limited by the cable connected to the vehicle so that remote control is only used in some particular cases. Thanks to 5G, the remote control can be performed without the limitation of cables so that 5G accelerates the usage of remote control. In this case, the cameras are usually installed on the machines to collect the surrounding environment information \cite{Yoshida.2019, Dadhich.2016, Johansson.2018}. Since they typically need more than three cameras to get the information, and the transmission rate of WIFI is limited, they cannot install more cameras to create the depth information resulting in lower productivity even with the very best operators \cite{Dadhich.2018}. Considering Virtual reality technology will be adopted with 5G, the difficulty of the remote control will be dramatically reduced. Better than the earlier network technologies, 5G guarantees the efficiency and accuracy of the remote control. Another major expected application is self-working machines. Cooperating with deep learning-based image processing models \cite{Xiang.2020d}, the image can be further processed on the local cloud. The command can then be directly sent to the machines. To avoid the additional cost, many scientists point out we can use a smartphone as an intermediary to transmit information instead of installing additional equipment \cite{Xiang.2020c}. 

In the above scenario, there are three key technologies for remotely controlling or self-working construction machinery. The first one is the high-speed data transmission rate. In order to enable the AI or human to fully understand the situation in realtime, construction machinery will under the sight of HD cameras or wears the cameras for an operator to get the video streaming data collection. The transmission of HD video requires a large bandwidth to ensure the fluency and realtime transmission of video content. The second is the low delay in receiving information. The realtime issuance of interactive behavior between operators and controlled construction machinery requires the network to have low latency to ensure that the controller's command can be executed in realtime through actuators. The third is the rapid and convenient communication network deployment between the construction machinery and the operators. If a wired network is used between the construction machinery and the controller, although the network delay and bandwidth can be guaranteed to a certain extent, the cable makes the activity of the construction machinery limited. Moreover,  the rapid deployment of networks between construction machinery and controllers cannot be easily achieved. 
If a 4G-LTE wireless cellular network is used, due to the limitation of the transmission rate and delay of the 4G-LTE network, the bandwidth and delay of the existing wireless network may not stably meet some high-rate and low-delay scenarios. These technical bottlenecks make the remote control collaboration project encounter many difficulties in the industry's practical application. No wonder so far, it has not been able to achieve widespread development and deployment. The large bandwidth and low delay technologies of the 5G network can solve these technical bottlenecks. 5G is bringing new opportunities for the industrial development of remote-controlled construction machinery.

\section{Problem statement and goal}
In the previous study from Bermudez \cite{Bermudez.2017}, they tested the performance of the LTE network by the transmission of video data. Their article evaluated two protocols' behaviors, realtime messaging protocol (RTMP) and realtime streaming protocol (RTSP), in a 4G environment. Based on their results, we can say that the performance of LTE to transfer the HD video from the working side to the operator side in realtime is good but not fully satisfying. 

Also, the through of LTE is in a steady-state growth situation. That means, the simulation parameters of Bermudez \cite{Bermudez.2017} missed the extreme working critical condition. We still don't know whether the LTE network can always have an excellent performance in a more stringent remote-control situation or not. Therefore, the comparison between LTE and 5G for video transmission in construction scenarios is necessary. For remote control, the delay is always a significant indicator because it equals to the accuracy and reliability of the job and the safety of the controlled machine \cite{Yoshida.2019}. Inspired from this and to fill this research gap, we decide to choose the performance from one of the new 5G technologies, mmWave, to compare with the LTE network's performance for construction machinery in remote-control and self-working scenarios.  Meanwhile, we will give the simulation a more stringent critical environment. Under the goal of finding out whether 5G network is more suitable for remote control or self-working construction machinery than LTE or not and if so, how good it is, our research is not in existence already.

\section{Modelling}
In the scenarios shown in Fig. \ref{fig: remote control} or Fig. \ref{fig: self-working}, that our UEs, i.e., construction machines, are under the sight of HD cameras or with the HD cameras. Here we assume our construction machines and cameras are both connected with the base station and the operator. The operator will give the construction machines commands. Meanwhile, they will collect the video streaming data from cameras. In case that the cameras are stick to the machine, the operator will give the order and receive the video data simultaneously. Compare with the instruction from the operator, video streaming data will occupy a much larger bandwidth. Therefore, in our research, we will use video streaming as the media, which can verify the performance of both networks. Obviously, video streaming with different resolution occupies different network bandwidth.  Depending on the different resolution requirements of video streaming, different pressure will be applied to the network.

For our research, we will use $ns-3$ \cite{Henderson.2008, Perrone.2009} as our simulation tool. To perform LTE simulation, we directly call the LTE module inside $ns-3$ because there is already a complete set of simulation modules and processes in $ns-3$ for 4G \cite{.0824202014:43:54}. On the other hand, for the 5G network, since it is still quite novel, $ns-3$ has not yet developed an official simulation platform with all 5G modules. Fortunately, because $ns-3$ is an open-source platform, many professional network simulation players can contribute to this platform based on their requirements, such as rewriting the algorithm, adding patch packages, or doing other upgrades. Among them, we selected the model from Mezzavilla \cite{Mezzavilla.2018} to simulate the 5G mmWave performance. The following paragraphs will present some basic architecture details and model settings for both network models. Basic parameters are shown in table \ref{tab: LTE} and table \ref{tab: 5G}. 

\begin{table}[htbp]
	\centering
	\caption{LTE Network Parameters, from 3GPP TS-36101 \cite{.0829202022:01:22}}
	\label{tab: LTE}
	\begin{small}
		\begin{tabular}{c|c}
			\hline
			\hline
			Parameters	        &   Value	\\
			\hline
			Bandwidth	        &   25MHz			\\ 	

			Downlink Earfcn		& 	100			 	\\

			Uplink Earfcn		& 	18100		\\

			Scheduler		    & 	\textit{PfFfMacScheduler}		\\
			\hline
			\hline
		\end{tabular}  
	\end{small}
\end{table}

\begin{table}[htbp]
	\centering
	\caption{5G Network Parameters, From 3GPP TS-38101 \cite{.0829202022:00:45}}
	\label{tab: 5G}
	\begin{small}
		\begin{tabular}{c|c}
			\hline
			\hline
			Parameters          &   Value        \\
			\hline
			Band	        &   n257			\\ 	
			
			Downlink NR-ARFCN		& 	2054167 – 2104165				\\
			
			Uplink NR-ARFCN		&   2054167 – 2104165			\\
			
			Scheduler		    & 	\textit{MmWaveMacScheduler}		\\
			\hline
			\hline
		\end{tabular}  
	\end{small}
\end{table}

\subsection{Model Parameters}

\subsubsection{Propagation Model}

For LTE, we use  \textit{FriisPropagationLossModel} \cite{Assyadzily.2014}. Given an unobstructed visual path between the transmitter and receiver, the free-space propagation model can predict the strength of the received signal. According to Friis \cite{Friis.1946}, the received signal strength can be described as, 
\begin{equation}
P_r(d)=\frac{P_tG_tG_r\lambda^2 }{(4\pi)^2d^2L} \label{eq1}
\end{equation}
where $P_r(d)$ is defined as received signal power, $P_t$ is transmit power, $G_t$ is transmit antenna gain, $G_r$ is receive antenna gain, $\lambda$ is wavelength(m), $d$ is the distance, and $L$ is the system loss.

As for 5G, we use  \textit{MmWavePropagationLossModel} \cite{3GPP., MacCartney.201312920131213}. This mmWave model presents two kinds of path loss models. The first one is the one that we used, which is in a statistical characteristic of the Line of Sight (LOS) state. The other one is \textit{BuildingsObstaclePropagationLossModel} \cite{.0829202022:25:39}, adding the obstacle between the gNB and the UE. Further path-loss models of mmWave can be found in \cite{3GPP.}.

\subsubsection{Transmission Control Protocol/Internet Protocol (TCP/IP)}

The network transmission adopts the TCP/IP protocol. The core protocols of the TCP/IP protocol are the transport layer protocol (TCP and UDP) and the network layer protocol (IP), which are usually implemented in the kernel of the operating system. Because the purpose of TCP is to achieve reliable data transmission, it has a set of handshake mechanism, send - confirmation, timeout - resend \cite{Wiebelitz.2009}.  In the case of video streaming, the network spending of TCP transmission is too large, thus impairing image quality and latency. Therefore, the UDP transmission method is preferred for realtime live streaming \cite{Madhuri.2016, Sinky.2015}.

\subsubsection{Hybrid Automatic Repeat Request (HARQ)}

For LTE and 5G, they both have two levels of retransmission mechanisms: HARQ at the MAC layer and ARQ at the Radio Link Control (RLC) layer \cite{Kim.2012, Anand.2018}. For 4G, the retransmission of lost or erroneous data is mainly handled by the HARQ mechanism of the MAC layer and supplemented by the ARQ of the RLC. The HARQ mechanism of the MAC layer can provide fast retransmission, and the ARQ mechanism of the RLC layer can provide reliable data transmission. In contrast, for 5G, the uplink HARQ mechanism is the same as the downlink, and both are asynchronous HARQ. There will be two kinds of changes \cite{Yeo.2017}. First, the scheduling timing is more flexible, especially in TDD mode, resulting in more resource allocation flexibility. Second, the pressure of data buffering will increase. Unlike LTE's uplink synchronous HARQ, asynchronous HARQ may have a longer retransmission interval. During this time, the UE must buffer the unACKed data, which will increase the buffering pressure.

\subsubsection{Scenarios}

We setup three scenarios for both network environments. In the first scenario, we choose 2Mbps as the video streaming volume. 2Mpbs is nearly the level of 720P video streaming bandwidth requirement \cite{Summerson.2018115}. Then we are changing the UE number from 2 to 20. In the second scenario, we set the UE number as a constant condition. By changing the data volume to realize new scenario, from 1Mbps to 8Mbps, which includes the bandwidth requirement of 720P(3Mbps), 1080P(5Mbps), and 3D 1080P(6Mbps) videos \cite{Bode.2020}, we explore the network performance with a varying resolution of the video. At last, we let UEs move to acquire the knowledge of how mobility condition affects the networks. For scenarios 1 and 2, our mobile machines will be under the sight of those high-definition cameras. Those cameras will collect the working video data and transfer it to the operator. The UEs in scenario three will be cameras installed on mobile machines. Here they will change their position together with the construction machinery as collecting the video streaming.

\begin{table}[htbp]
	\centering
	\caption{Network Scenario 1}
	\label{tab:Network Scenario 1}
	\begin{small}
		\begin{tabular}{c|c}
			Data Volume(Mbps)	        &   2			\\ 	
			\hline
			UE Number	    	& 	2, 4, 6, 8, 10, 12, 14, 16, 18, 20				\\

		\end{tabular}  
	\end{small}
\end{table}

\begin{table}[htbp]
	\centering
	\caption{Network Scenario 2}
	\label{tab:Network Scenario 2}
	\begin{small}
		\begin{tabular}{c|c}
			UE Number	        &   8			\\ 	
			\hline
			Data Volume (Mbps)	    	& 	1, 2, 3, 4, 5, 6, 7, 8				\\
			
		\end{tabular}  
	\end{small}
\end{table}

\begin{table}[htbp]
	\centering
	\caption{Network Scenario 3}
	\label{tab:Network Scenario 3}
	\begin{small}
		\begin{tabular}{c|c}
			Data Volume(Mbps)	        &   2		\\ 	
			\hline
			UE Number	    	& 	8				\\
			\hline
			UE Distance(m)            &   20, 40, 60, 80, 100, 120, 140, 160, 180, 200
		
		\end{tabular}  
	\end{small}
\end{table}

\section{Simulation Results}

This section presents the results of our simulated network scenarios in terms of throughput, packet loss rate, and delay. As for both network environments, we performed the simulation repeatedly and got the average value to improve accuracy. 

\begin{figure}[h!]
\centerline{\includegraphics[width=3.0in]{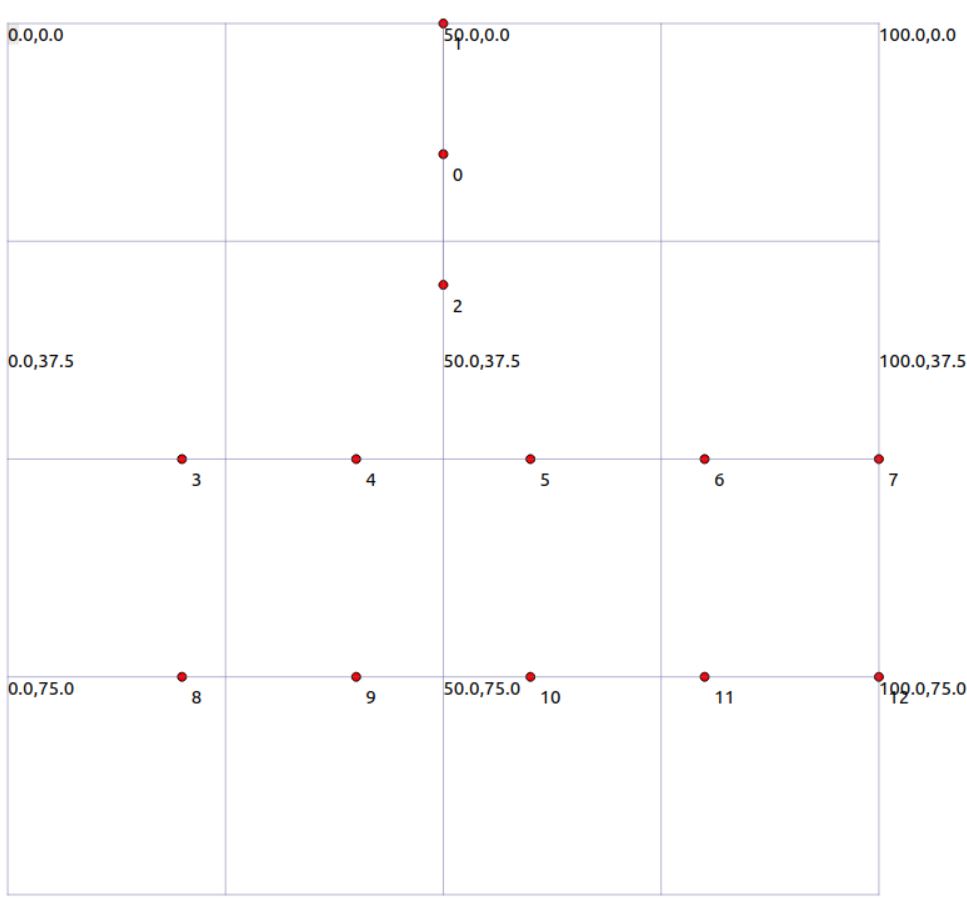}}
\caption{Network topology.}
\label{fig: network topology}
\end{figure}

\begin{figure*}[!t]
         \newcommand{\w}{0.25}
         \centering 
         \subfloat[Average throughput with increasing UE]{
             \includegraphics[width=\w\textwidth]{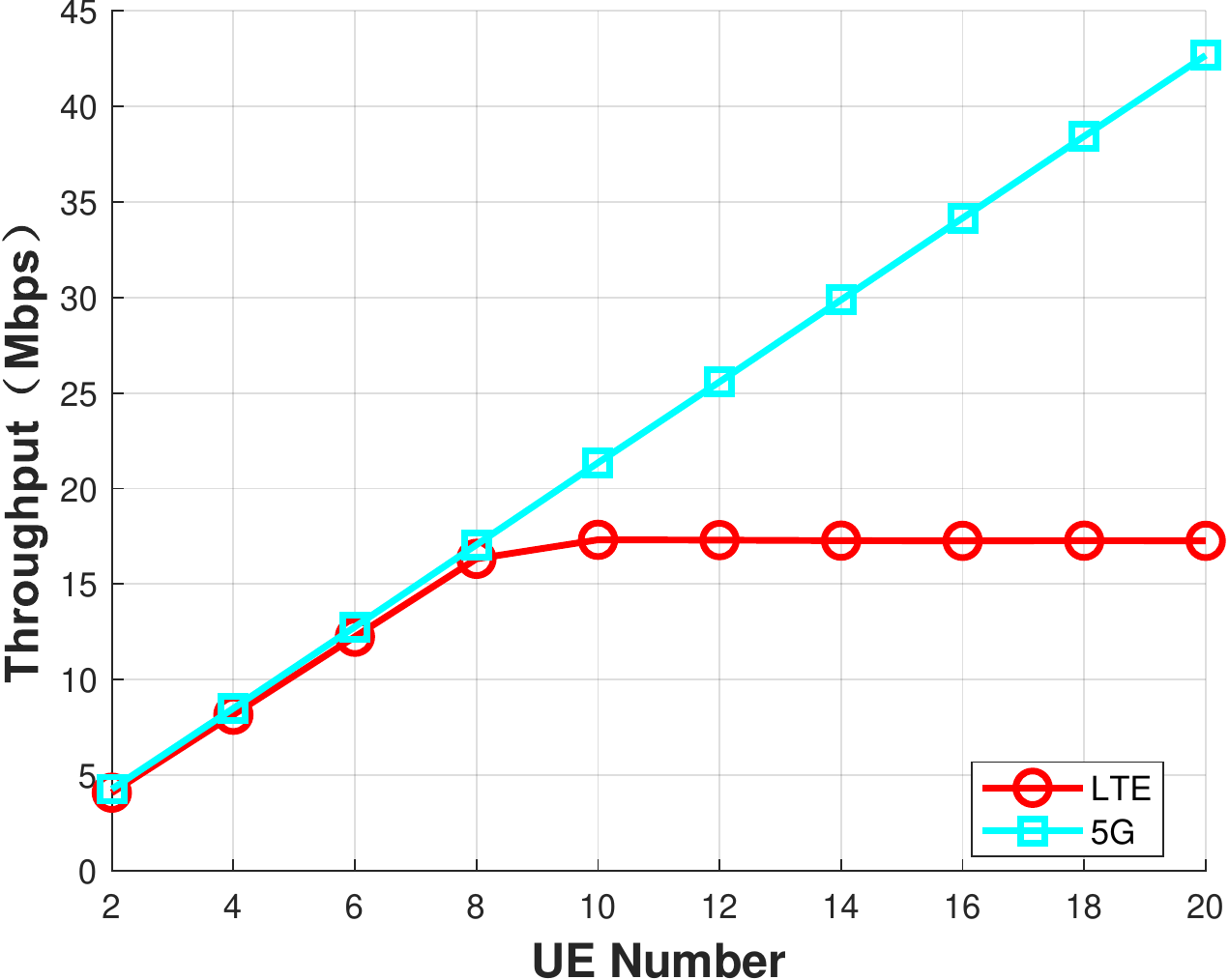}}
             \hfil 
         \subfloat[Packet loss rate with increasing UE]{
             \includegraphics[width=\w\textwidth]{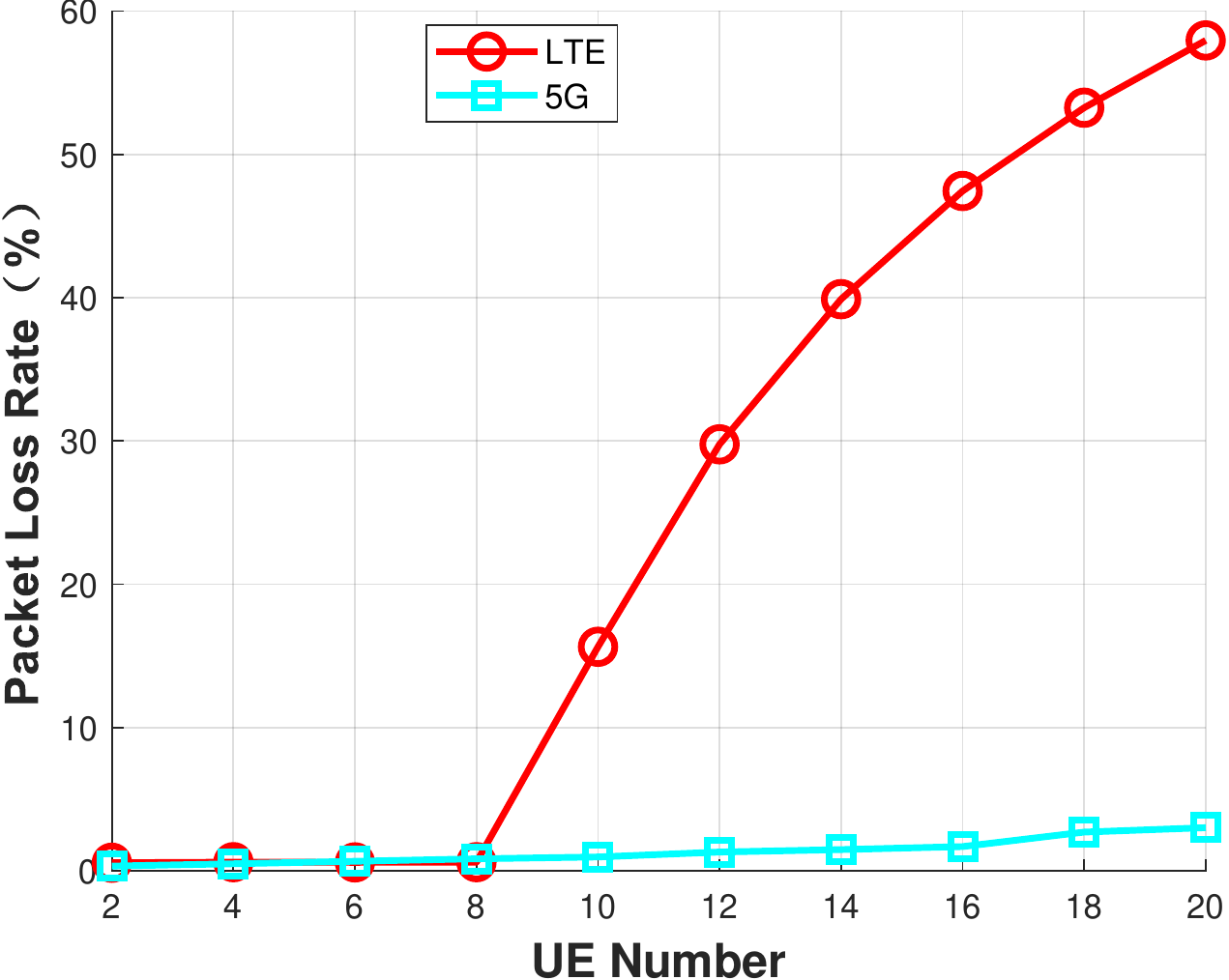}} 
             \hfil
         \subfloat[Average delay with increasing UE]{
             \includegraphics[width=\w\textwidth]{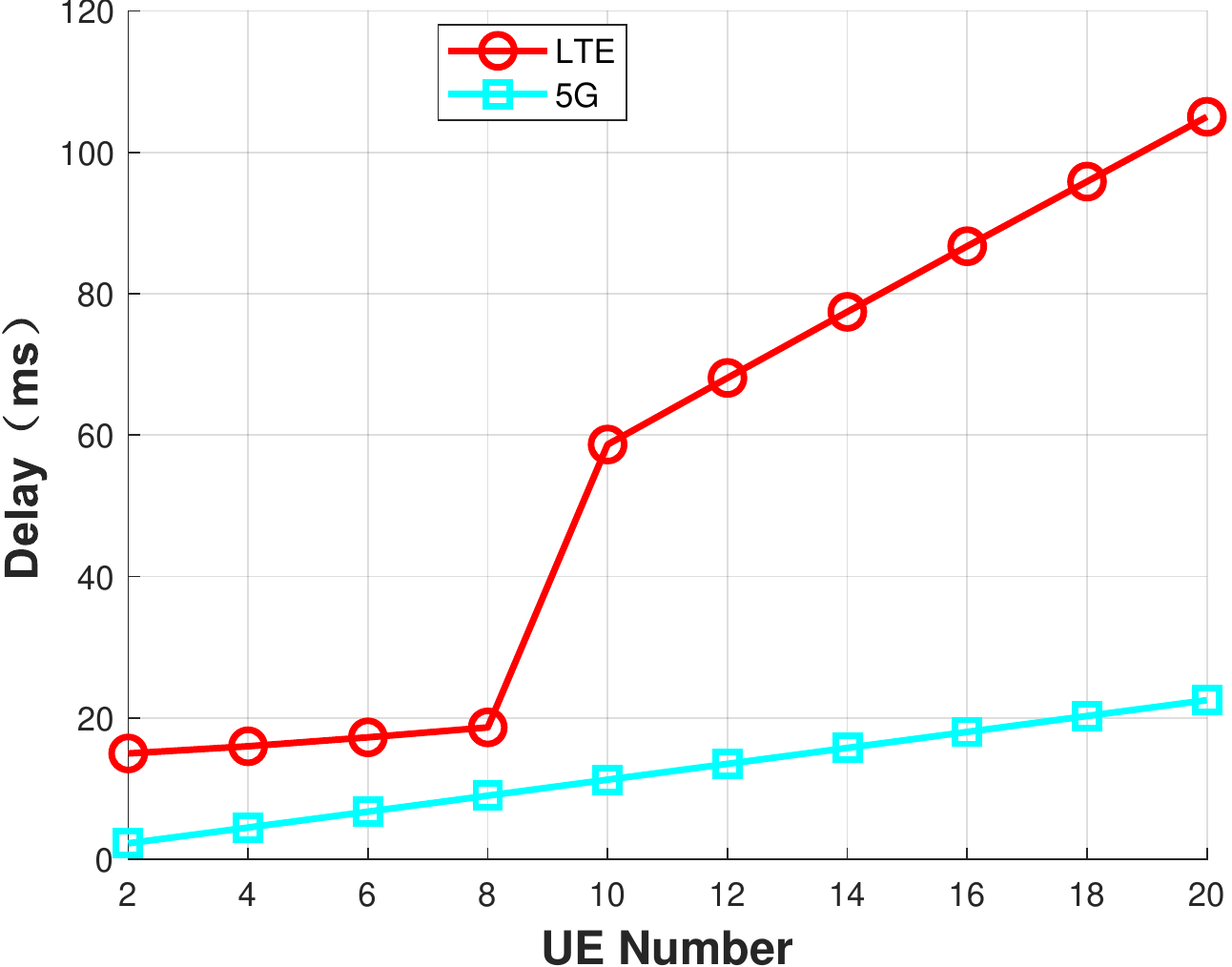}} 
             \hfil
         \subfloat[Average throughput with increasing Data Volume]{
             \includegraphics[width=\w\textwidth]{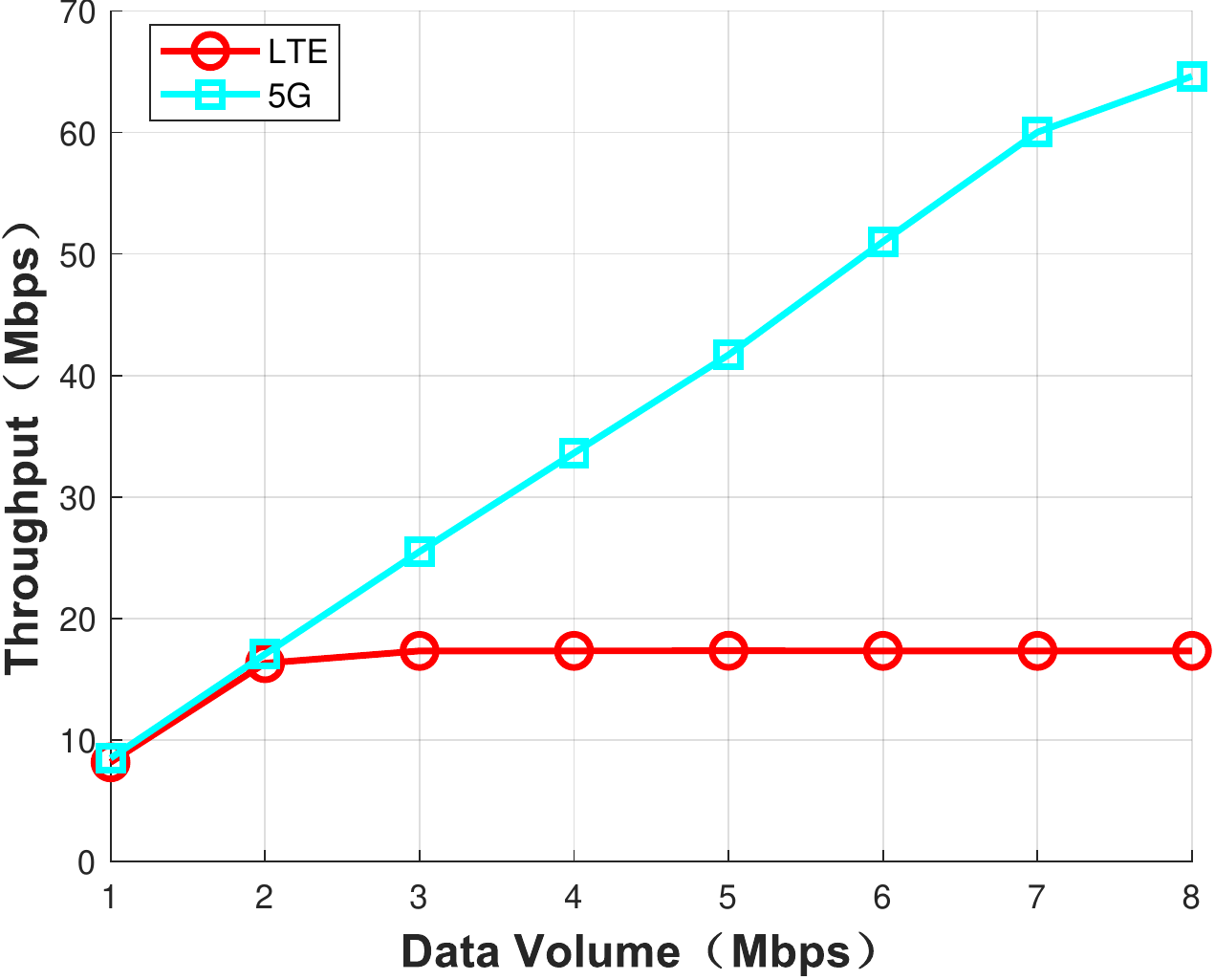}}
             \hfil 
         \subfloat[Packet loss rate with increasing Data Volume]{
            \includegraphics[width=\w\textwidth]{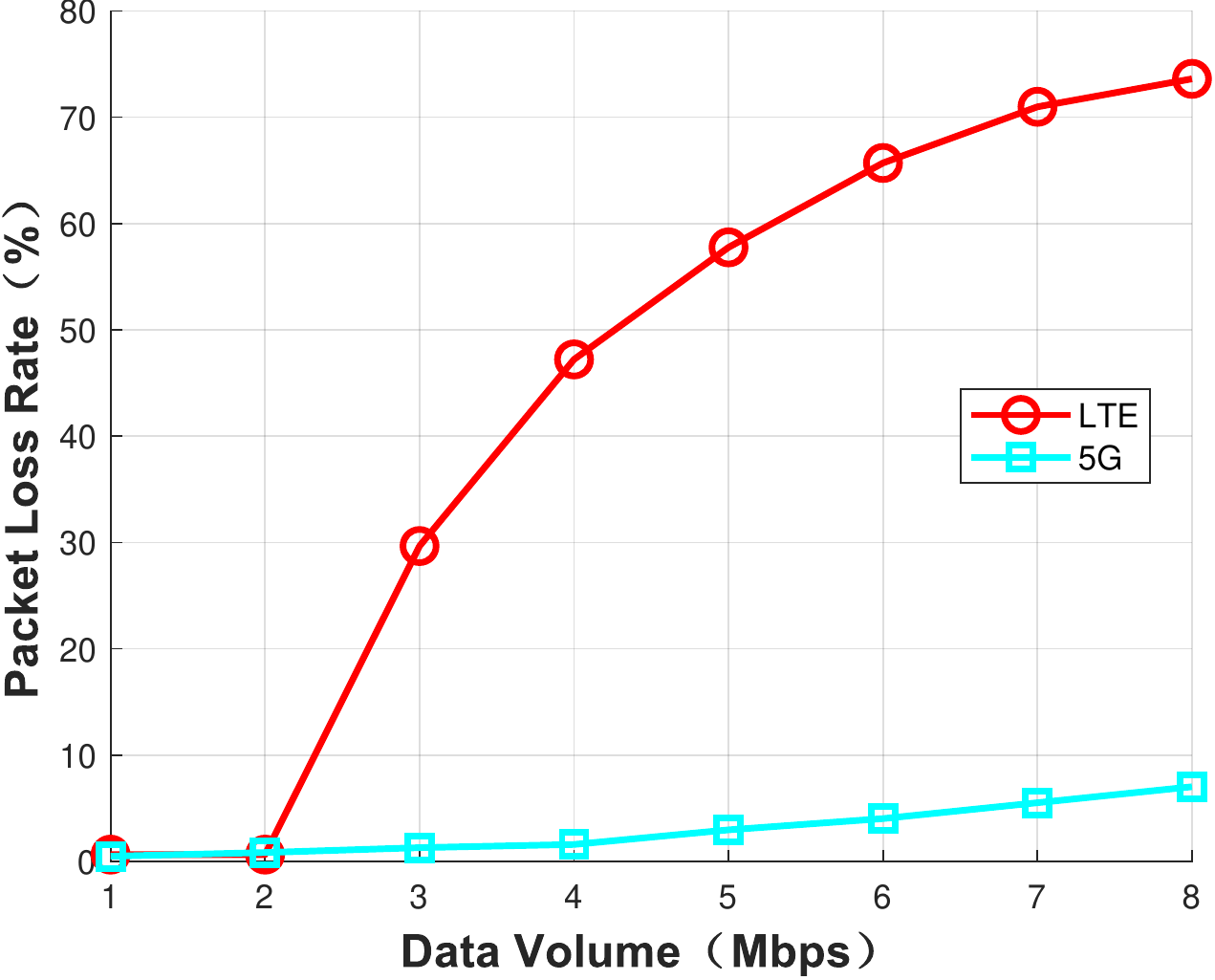}} 
             \hfil
        \subfloat[Average delay with increasing Data Volume]{
             \includegraphics[width=\w\textwidth]{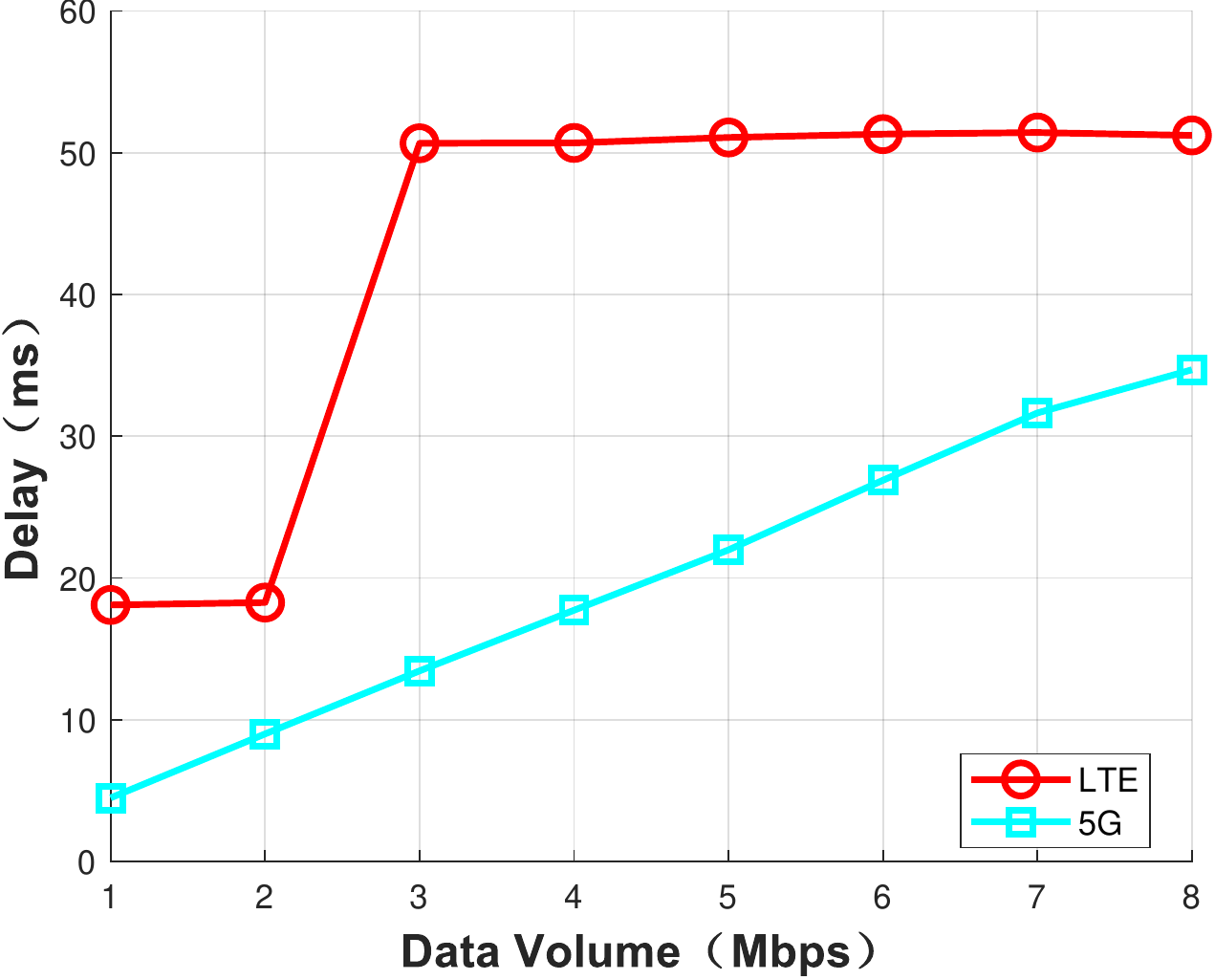}} 
             \hfil
         \caption{Simulation results of scenario 1 and scenario 2} \label{fig:Simulation results of scenario 1 and scenario 2}
\end{figure*} 

\begin{figure*}[!t]
         \newcommand{\w}{0.25}
         \centering 
         \subfloat[Average throughput with increasing Velocity]{
             \includegraphics[width=\w\textwidth]{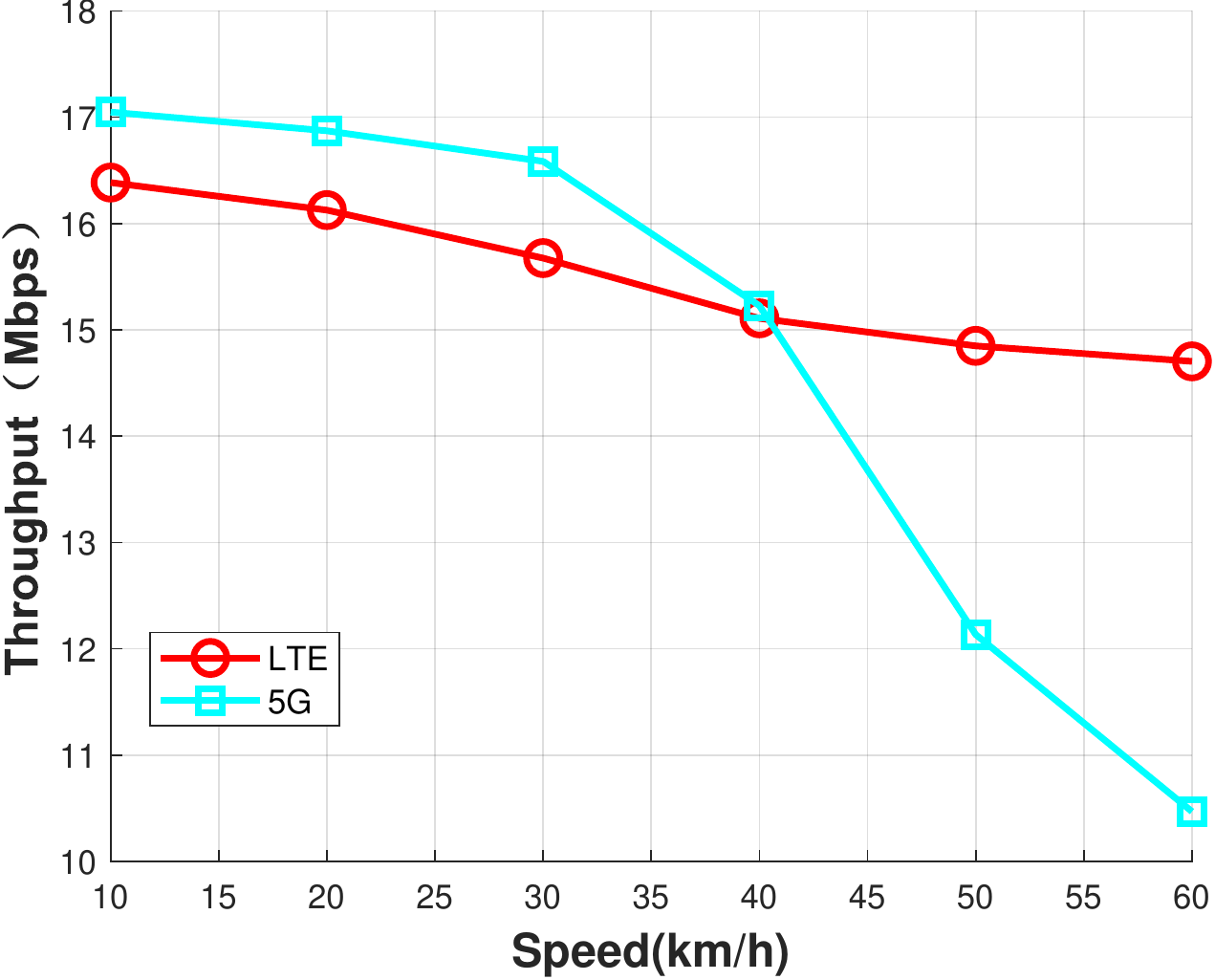}}
             \hfil 
         \subfloat[Packet loss rate with increasing Velocity]{
             \includegraphics[width=\w\textwidth]{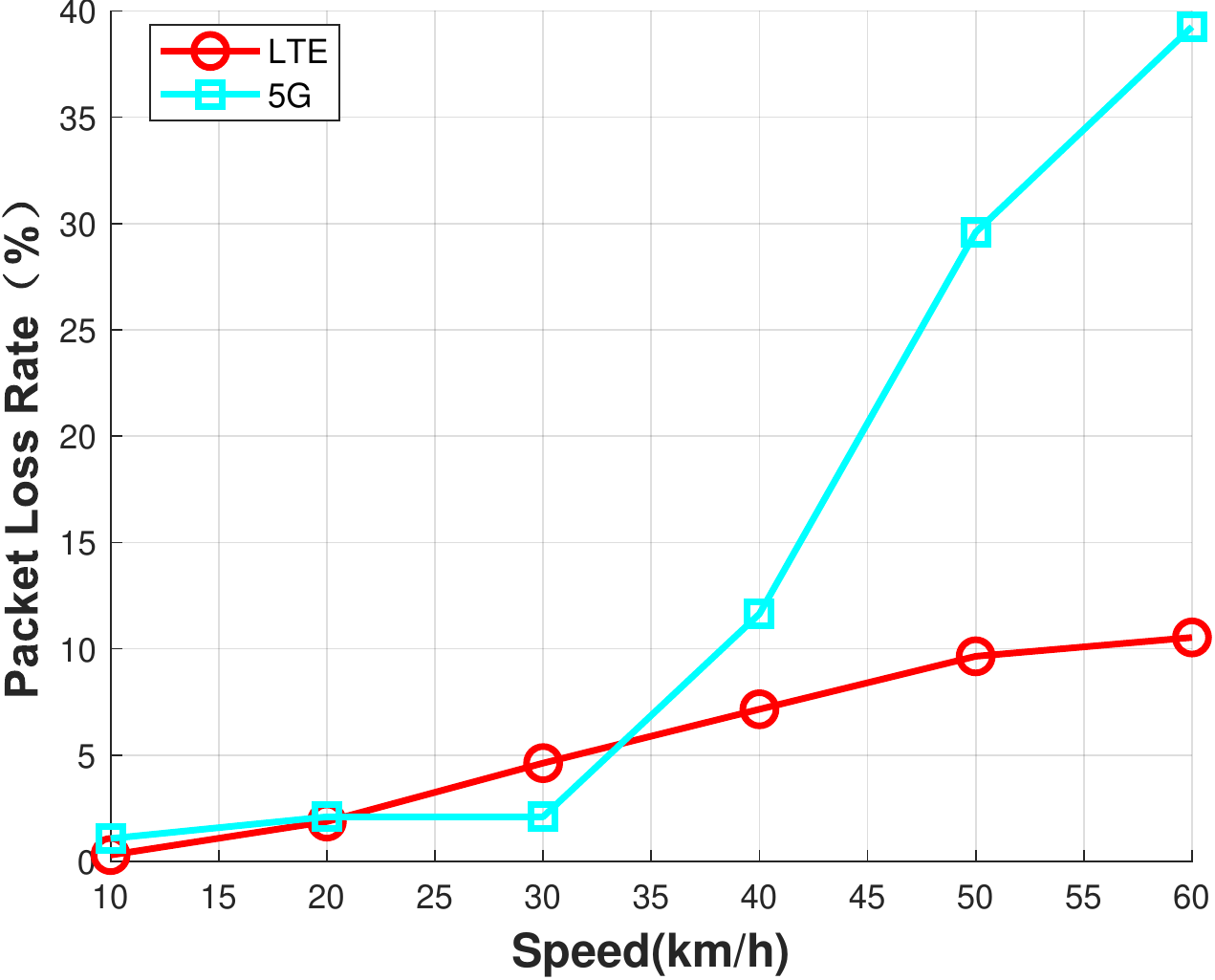}} 
             \hfil
         \subfloat[Average delay with increasing Velocity]{
             \includegraphics[width=\w\textwidth]{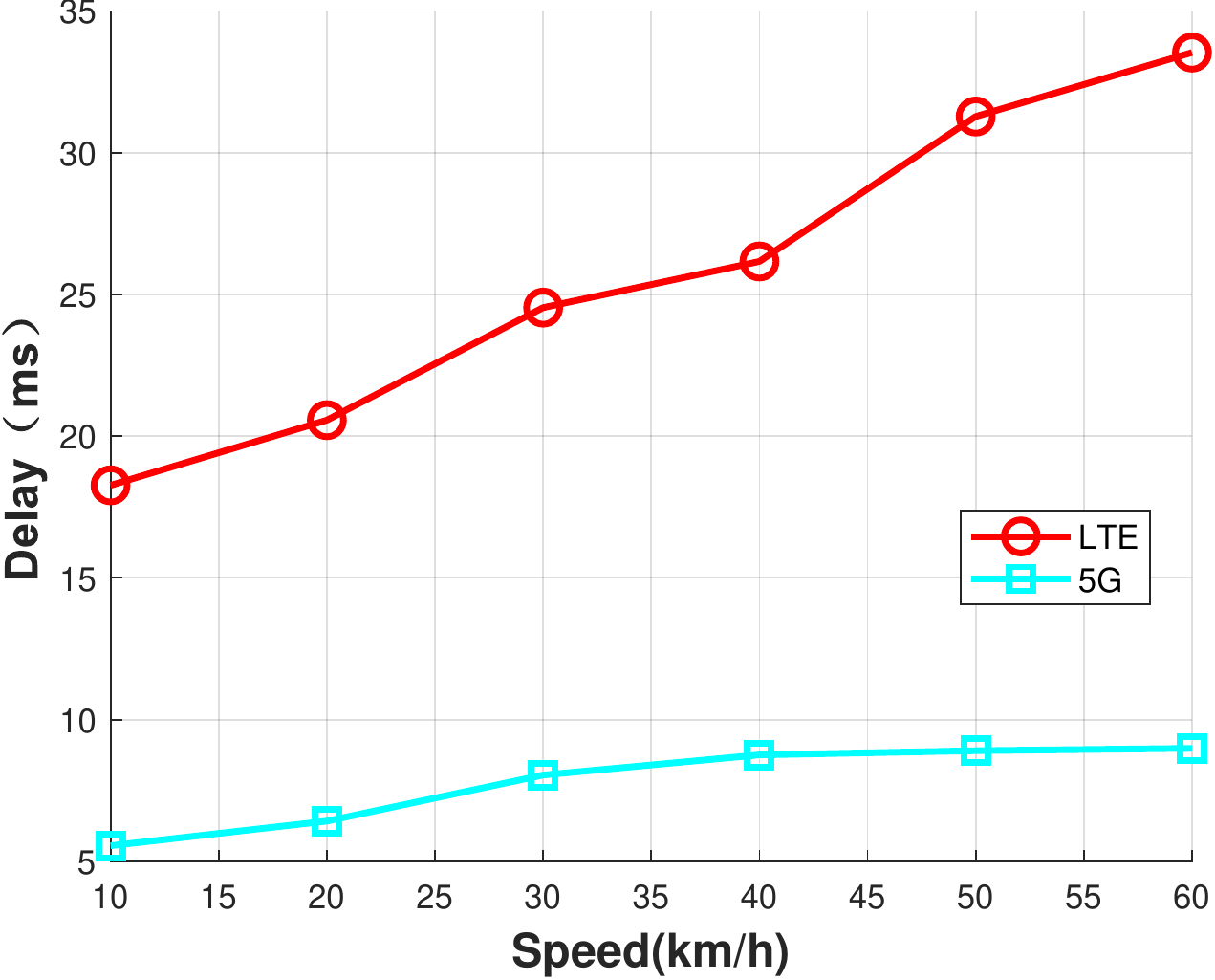}} 
             \hfil
         
         \caption{Simulation results of scenario 3} \label{fig:Simulation results of scenario 3}
\end{figure*}

The network topology is shown in Fig. \ref{fig: network topology}. From nodes 3 to node 12 represent a set of remote devices, i.e., cameras, and the transmission data represents the video data sent by the camera avatar, which is finally sent to the user terminal (node 1) through eNodeB or NR (node 2) and Evolved packet core (EPC) (node 0).

With the increase in the number of UE, the throughput simulation results are shown in Fig. \ref{fig:Simulation results of scenario 1 and scenario 2}(a). In the beginning, the throughput of LTE and 5G networks has increased rapidly, and the throughput matches the total data volume, which means both of them can complete the transmission of the video streaming task. As the number of UEs further increases, the 5G network can still transmit video service data better; however, the LTE network cannot provide enough transmission capacity for video service data, reaching a state of business saturation. It can be observed that the throughput remains basically unchanged as the UE number grows, about 17Mbps. In Fig. \ref{fig:Simulation results of scenario 1 and scenario 2}(b) simulation results of the packet loss rate as the UE number increase are shown. When the UE number is small, both LTE and 5G networks can keep the packet loss rate approximately equal to 0, i.e., almost no packet loss occurs. If the UE number increases, the 5G network can still maintain the network with an almost low packet loss rate. Still, the LTE network will have more packet loss due to its network resource constraints. It must discard the video service's data packets, causing the transmitted video to lose frames, freeze or completely lose the result of the video image, which will seriously affect the operator's performance of the construction machinery. Besides that, high latency will make a video to be out of sync. In these cases, the operator cannot grasp the on-site working environment in realtime, resulting in the operator to make wrong judgments about the working environment, which is very dangerous for the work task and the construction machinery. The average delay of the 5G network is lower than that of the LTE network, as shown in Fig. \ref{fig:Simulation results of scenario 1 and scenario 2}(c). This is because the 5G network can provide larger network bandwidth, increase network transmission speed, and reduce data packet delay. If the UE number is small, the average delay of the LTE network is about twice that of the 5G network; however, when the number of users is large, the average delay of the LTE network is much higher than that of the 5G network. At this time, the LTE network cannot guarantee the video streaming service.

In the second simulation scenario, the number of UE number is fixed to 8, and the video service data is increased from 1Mbps to 8Mbps. The simulation result of throughput with increasing video service rate is shown in Fig. \ref{fig:Simulation results of scenario 1 and scenario 2}(d). When the video service rates are 1Mbps and 2Mbps, the throughput of the LTE network and the 5G network can meet video streaming services' requirements. However, when the video service rate exceeds 3Mbps, the throughput of the LTE network does not continue to increase, and the throughput of the 5G network still increases with the video service rate, which can guarantee the transmission of the video service. The simulation results of the packet loss rate are demonstrated in Fig. \ref{fig:Simulation results of scenario 1 and scenario 2}(e), where we can see that the 5G network has been able to maintain the packet loss rate at a low level. However, severe packet loss will occur for LTE networks when a higher video service rate is required. In case that the video service rate is 5Mbps, the packet loss rate of LTE exceeds 50\%. The average delay of video services is presented in Fig. \ref{fig:Simulation results of scenario 1 and scenario 2}(f). 5G network continues to increase with the increase of data volume, and they are all maintained at a low level, even when the video service rate is 5Mbps, the average delay still does not exceed 25ms. The video service average delay of the LTE network is significantly higher than that of the 5G network. In short, as the video service rate goes higher, the improvement with 5G will be more significant.

In the third scenario, we want to simulate the case that construction machines carry the cameras with them when they change their positions. Here the video service data rate is 2Mbps, and the number of remote devices is still 8. We simulate the longest distance up to 200m since the longest propagation distance of mmWave is considered as 200m \cite{Seker.2018}. The simulation results of throughput with increasing speed are shown in Fig. \ref{fig:Simulation results of scenario 3}(a). Due to lower frequency bands, LTE network performances are affected only slightly with mobility. 
Also, when the UE velocity is lower than 40km/h, the throughput of the 5G network is still in a relatively stable decline stage. However, when the UE velocity exceeds 40km/h, the throughput of the 5G network drops dramatically, and thus the transmission of video services cannot be guaranteed at this time. Fig. \ref{fig:Simulation results of scenario 3}(b) presents, as the velocity increases, the packet loss rate is rising slowly for LTE networks. However, the 5G network will suffer a fast increasing packet loss rate when the UE moves faster than 30km/h. In Fig. \ref{fig:Simulation results of scenario 3}(c), both the delay of the LTE network and 5G network increase steadily with the growth of velocity. However, we can notice that the delay of the 5G network still much advantageous compare with LTE. 

To sum up, 5G mmWave has significant advantages in terms of throughput, packet loss, and latency if the UEs are fixed. Although one of the requirements for 5G is the capacity to deal with high mobility, the mmWave 5G may still have a problem if the beamforming technology, concretely, tracking algorithm, is not perfect. In contrast, since 4G uses a lower frequency band, this problem is not so apparent for 4G, which hints the suitability of using sub 6Ghz band 5G.

\section{Conclusion}
In this paper, we first indicate that 5G can be employed in the construction machines industry to improve the remote control operation and work as an essential component to achieve self-working construction machines. By taking the remote-control and self-working of construction machinery as the scenes and using video streaming transmission as the medium, we compared the LTE network's performance and the 5G mmWave network. Based on our research, we found that 5G has the capability to accomplish a better quality of live streaming so that both scenes can be significantly improved. Especially, 5G can let more cameras in the same network, indicating the possibility to acquire depth information from the video. Besides, since it is not difficult to let the machines always under the cameras' vision, we suggest letting the cameras unmoved avoid the shortcoming of mmWave. Otherwise, more robust beamforming, i.e., dynamic beamforming, algorithm is needed. 

Since we use video as the medium to test the performance of the two networks, future work shall refine video factors and explore how the structure of the different encoding video style will affect the networks. Besides, starting from the video phase, through the networks, and finally to the control operator, a simulation analysis of the entire link can be carried out to improve the content of this article. 
Moreover, as the 6G technology is on the way \cite{Letaief.2019,Peng.2016, Saad.2020}, we will explore the possibility to benefit the construction machine industry from 6G technology in our next paper.

\bibliography{Literature.bib}{}

\begin{thebibliography}{10}
\providecommand{\url}[1]{#1}
\csname url@samestyle\endcsname
\providecommand{\newblock}{\relax}
\providecommand{\bibinfo}[2]{#2}
\providecommand{\BIBentrySTDinterwordspacing}{\spaceskip=0pt\relax}
\providecommand{\BIBentryALTinterwordstretchfactor}{4}
\providecommand{\BIBentryALTinterwordspacing}{\spaceskip=\fontdimen2\font plus
\BIBentryALTinterwordstretchfactor\fontdimen3\font minus
  \fontdimen4\font\relax}
\providecommand{\BIBforeignlanguage}[2]{{%
\expandafter\ifx\csname l@#1\endcsname\relax
\typeout{** WARNING: IEEEtran.bst: No hyphenation pattern has been}%
\typeout{** loaded for the language `#1'. Using the pattern for}%
\typeout{** the default language instead.}%
\else
\language=\csname l@#1\endcsname
\fi
#2}}
\providecommand{\BIBdecl}{\relax}
\BIBdecl

\bibitem{Xiang.2020b}
Y.~Xiang, T.~Su, C.~Brach, X.~Liu, and M.~Geimer, ``Realtime estimation of ieee
  802.11p for mobile working machines communication respecting delay and packet
  loss,'' in \emph{IEEE Intelligent Vehicle Symposium 2020}, Las Vegas, USA,
  2020.

\bibitem{Henderson.2008}
T.~R. Henderson, M.~Lacage, G.~F. Riley, C.~Dowell, and J.~Kopena, Eds.,
  \emph{Network simulations with the ns-3 simulator}.\hskip 1em plus 0.5em
  minus 0.4em\relax New York, NY: ACM, 2008.

\bibitem{Xiang.2020e}
Y.~Xiang, Y.~Huang, Z.~Zhang, T.~Su, C.~Brach, S.~Mao, and M.~Geimer,
  ``Kit~moma~v2: towards instance segmentation of construction machines,''
  \emph{arXiv}, 2020.

\bibitem{Ghosh.2010}
A.~Ghosh, R.~Ratasuk, B.~Mondal, N.~Mangalvedhe, and T.~Thomas, ``Lte-advanced:
  next-generation wireless broadband technology,'' \emph{IEEE Wireless
  Communications}, vol.~17, no.~3, pp. 10--22, 2010.

\bibitem{Chen.2014}
S.~Chen and J.~Zhao, ``The requirements, challenges, and technologies for 5g of
  terrestrial mobile telecommunication,'' \emph{IEEE Communications Magazine},
  vol.~52, no.~5, pp. 36--43, 2014.

\bibitem{Gupta.2015}
A.~Gupta and R.~K. Jha, ``A survey of 5g network: Architecture and emerging
  technologies,'' \emph{IEEE Access}, vol.~3, pp. 1206--1232, 2015.

\bibitem{Lauridsen.2017}
M.~Lauridsen, L.~C. Gimenez, I.~Rodriguez, T.~B. Sorensen, and P.~Mogensen,
  ``From lte to 5g for connected mobility,'' \emph{IEEE Communications
  Magazine}, vol.~55, no.~3, pp. 156--162, 2017.

\bibitem{Tahir.2019}
M.~N. Tahir, K.~Maenpaa, and T.~Sukuvaara, ``Evolving wireless vehicular
  communication system level comparison and analysis of 802, 11 p, 4g 5g,'' in
  \emph{2nd International Conference on Communication, Computing and Digital
  systems (C-CODE)}.\hskip 1em plus 0.5em minus 0.4em\relax IEEE, 2019, pp.
  48--52.

\bibitem{Vinel.2017}
A.~Vinel, J.~Breu, T.~H. Luan, and H.~Hu, ``Emerging technology for 5g-enabled
  vehicular networks,'' \emph{IEEE Wireless Communications}, vol.~24, no.~6,
  p.~12, 2017.

\bibitem{Shafi.2017}
M.~Shafi, A.~F. Molisch, P.~J. Smith, T.~Haustein, P.~Zhu, P.~de~Silva,
  F.~Tufvesson, A.~Benjebbour, and G.~Wunder, ``5g: A tutorial overview of
  standards, trials, challenges, deployment, and practice,'' \emph{IEEE Journal
  on Selected Areas in Communications}, vol.~35, no.~6, pp. 1201--1221, 2017.

\bibitem{.0829202022:00:45}
\BIBentryALTinterwordspacing
3GPP, ``3gpp specification series: 38series,'' 08/29/2020 22:00:45. [Online].
  Available: \url{https://www.3gpp.org/DynaReport/38-series.htm}
\BIBentrySTDinterwordspacing

\bibitem{Deng.2017}
J.~Deng, O.~Tirkkonen, R.~Freij-Hollanti, T.~Chen, and N.~Nikaein, ``Resource
  allocation and interference management for opportunistic relaying in
  integrated mmwave/sub-6 ghz 5g networks,'' \emph{IEEE Communications
  Magazine}, vol.~55, no.~6, pp. 94--101, 2017.

\bibitem{Li.2018}
Y.~Li, C.~Sim, Y.~Luo, and G.~Yang, ``12-port 5g massive mimo antenna array in
  sub-6ghz mobile handset for lte bands 42/43/46 applications,'' \emph{IEEE
  Access}, vol.~6, pp. 344--354, 2018.

\bibitem{Han.2015}
S.~Han, C.~I, Z.~Xu, and C.~Rowell, ``Large-scale antenna systems with hybrid
  analog and digital beamforming for millimeter wave 5g,'' \emph{IEEE
  Communications Magazine}, vol.~53, no.~1, pp. 186--194, 2015.

\bibitem{Campos.2017}
J.~Campos, ``Understanding the 5g nr physical layer.''

\bibitem{Mohyeldin.2016}
E.~Mohyeldin, ``Minimum technical performance requirements for imt-2020 radio
  interface (s),'' in \emph{ITU-R Workshop on IMT-2020 Terrestrial Radio
  Interfaces}, 2016, pp. 1--12.

\bibitem{Ansari.2018}
R.~I. Ansari, C.~Chrysostomou, S.~A. Hassan, M.~Guizani, S.~Mumtaz,
  J.~Rodriguez, and {Rodrigues, J. J. P. C.}, ``5g d2d networks: Techniques,
  challenges, and future prospects,'' \emph{IEEE Systems Journal}, vol.~12,
  no.~4, pp. 3970--3984, 2018.

\bibitem{Roh.2014}
W.~Roh, J.~Y. Seol, J.~Park, B.~Lee, J.~Lee, Y.~Kim, J.~Cho, K.~Cheun, and
  F.~Aryanfar, ``Millimeter-wave beamforming as an enabling technology for 5g
  cellular communications: theoretical feasibility and prototype results,''
  \emph{IEEE Communications Magazine}, vol.~52, no.~2, pp. 106--113, 2014.

\bibitem{Yoshida.2019}
H.~Yoshida, T.~Yoshimoto, T.~Innami, K.~Ohashi, H.~Furuya, and N.~Mori,
  ``Improving efficiency of remote construction by using adaptive video
  streaming,'' in \emph{2019 16th IEEE Annual Consumer Communications {\&}
  Networking Conference (CCNC), IEEE}, 2019.

\bibitem{Dadhich.2016}
S.~Dadhich, U.~Bodin, and U.~Andersson, ``Key challenges in automation of
  earth-moving machines,'' \emph{Automation in Construction}, vol.~68, pp.
  212--222, 2016.

\bibitem{Johansson.2018}
I.~Johansson, S.~Dadhich, U.~Bodin, and T.~J{\"o}nsson, ``Adaptive video with
  scream over lte for remote-operated working machines,'' \emph{Wireless
  Communications and Mobile Computing}, vol. 2018, pp. 1--10, 2018.

\bibitem{Dadhich.2018}
S.~Dadhich, U.~Bodin, F.~Sandin, and U.~Andersson, ``From tele-remote operation
  to semi-automated wheel-loader,'' \emph{International Journal of Electrical
  and Electronic Engineering and Telecommunications}, vol.~7, no.~4, pp.
  178--182, 2018.

\bibitem{Xiang.2020d}
\BIBentryALTinterwordspacing
Y.~Xiang, Y.~Wang, T.~Su, R.~Li, C.~Brach, S.~Mao, and M.~Geimer, ``Kit moma: A
  mobile machines dataset,'' \emph{arXiv}, 2020. [Online]. Available:
  \url{arXiv:2007.04198}
\BIBentrySTDinterwordspacing

\bibitem{Xiang.2020c}
\BIBentryALTinterwordspacing
Y.~Xiang, T.~Tang, T.~Su, C.~Brach, L.~Liu, S.~Mao, and M.~Geimer, ``Fast
  crdnn: Towards on site training of mobile construction machines,''
  \emph{arXiv}, 2020. [Online]. Available:
  \url{https://arxiv.org/pdf/2006.03169.pdf}
\BIBentrySTDinterwordspacing

\bibitem{Bermudez.2017}
\BIBentryALTinterwordspacing
H.~F. Bermudez, R.~Sanchez-Iborra, and J.~L. Arciniegas, ``Performance
  validation of ns3-lte emulation for live video streaming under qos
  parameters. in , networking and communications (wimob),'' in \emph{2017 IEEE
  13th International Conference on Wireless and Mobile Computing Networking and
  Communications (WiMob)}, 2017, pp. 300--307. [Online]. Available:
  \url{http://ieeexplore.ieee.org/servlet/opac?punumber=8106925}
\BIBentrySTDinterwordspacing

\bibitem{Perrone.2009}
L.~F. Perrone, C.~Cicconetti, G.~Stea, and B.~C. Ward, ``On the automation of
  computer network simulators,'' in \emph{Proceedings of the 2nd International
  Conference on Simulation Tools and Techniques}, 2009, pp. 1--10.

\bibitem{.0824202014:43:54}
\BIBentryALTinterwordspacing
ns3, ``Lte module --- model library,'' 08/24/2020 14:43:54. [Online].
  Available: \url{https://www.nsnam.org/docs/models/html/lte.html}
\BIBentrySTDinterwordspacing

\bibitem{Mezzavilla.2018}
M.~Mezzavilla, M.~Zhang, M.~Polese, R.~Ford, S.~Dutta, S.~Rangan, and M.~Zorzi,
  ``End-to-end simulation of 5g mmwave networks,'' \emph{IEEE Communications
  Surveys {\&} Tutorials}, vol.~20, no.~3, pp. 2237--2263, 2018.

\bibitem{.0829202022:01:22}
\BIBentryALTinterwordspacing
3GPP, ``3gpp specification series: 36series,'' 08/29/2020 22:01:22. [Online].
  Available: \url{https://www.3gpp.org/DynaReport/36-series.htm}
\BIBentrySTDinterwordspacing

\bibitem{Assyadzily.2014}
M.~Assyadzily, A.~Suhartomo, and A.~Silitonga, Eds., \emph{Evaluation of
  X2-handover performance based on RSRP measurement with Friis path loss using
  network simulator version 3 (NS-3)}.\hskip 1em plus 0.5em minus 0.4em\relax
  {2014 2nd International Conference on Information and Communication
  Technology (ICoICT), IEEE}, 2014.

\bibitem{Friis.1946}
H.~T. Friis, ``A note on a simple transmission formula,'' in \emph{Proceedings
  of the IRE}, vol. 34(5), 1946, pp. 254--256.

\bibitem{3GPP.}
3GPP, ``Study on channel model for frequency spectrum above 6 ghz.''

\bibitem{MacCartney.201312920131213}
G.~R. MacCartney, J.~Zhang, S.~Nie, and T.~S. Rappaport, ``Path loss models for
  5g millimeter wave propagation channels in urban microcells,'' in \emph{2013
  IEEE Global Communications Conference (GLOBECOM)}.\hskip 1em plus 0.5em minus
  0.4em\relax IEEE, 2013/12/9 - 2013/12/13, pp. 3948--3953.

\bibitem{.0829202022:25:39}
\BIBentryALTinterwordspacing
ns3, ``ns-3: ns-3 documentation,'' 08/29/2020 22:25:39. [Online]. Available:
  \url{https://www.nsnam.org/doxygen/}
\BIBentrySTDinterwordspacing

\bibitem{Wiebelitz.2009}
J.~Wiebelitz, C.~Kunz, S.~Piger, and C.~Grimm, ``Tcp-authn: Tcp inline
  authentication to enhance network security in grid environments,'' in
  \emph{2009 Eighth International Symposium on Parallel and Distributed
  Computing, IEEE}, 2009, pp. 237--240.

\bibitem{Madhuri.2016}
D.~Madhuri and P.~C. Reddy, ``Performance comparison of tcp, udp and sctp in a
  wired network,'' in \emph{2016 International Conference on Communication and
  Electronics Systems (ICCES), IEEE}, 2016, pp. 1--6.

\bibitem{Sinky.2015}
M.~Sinky, A.~Dhamodaran, B.~Lee, and J.~Zhao, ``Analysis of h.264 bitstream
  prioritization for dual tcp/udp streaming of hd video over wlans,'' in
  \emph{2015 12th Annual IEEE Consumer Communications and Networking Conference
  (CCNC), IEEE}, 2015, pp. 576--581.

\bibitem{Kim.2012}
M.~Kim, S.~Kim, and Y.~Lim, ``An implementation of downlink asynchronous harq
  for lte tdd system,'' in \emph{2012 IEEE Radio and Wireless Symposium}.\hskip
  1em plus 0.5em minus 0.4em\relax IEEE, 2012, pp. 271--274.

\bibitem{Anand.2018}
A.~Anand and G.~de~Veciana, Eds., \emph{Resource Allocation and HARQ
  Optimization for URLLC Traffic in 5G Wireless Networks}, vol.~36.\hskip 1em
  plus 0.5em minus 0.4em\relax {IEEE Journal on Selected Areas in
  Communications, IEEE}, 2018.

\bibitem{Yeo.2017}
J.~Yeo, S.~Park, J.~Oh, Y.~Kim, and J.~Lee, ``Partial retransmission scheme for
  harq enhancement in 5g wireless communications,'' in \emph{2017 IEEE Globecom
  Workshops (GC Wkshps), IEEE}, 2017.

\bibitem{Summerson.2018115}
\BIBentryALTinterwordspacing
C.~Summerson, ``How much data does netflix use?'' \emph{How-To Geek},
  2018/1/15. [Online]. Available:
  \url{https://www.howtogeek.com/338983/how-much-data-does-netflix-use/}
\BIBentrySTDinterwordspacing

\bibitem{Bode.2020}
\BIBentryALTinterwordspacing
K.~Bode, ``Netflix starts offering 'super hd' and 3d streams - but only through
  isps that use their cdn,'' 2020. [Online]. Available:
  \url{http://www.dslreports.com/shownews/Netflix-Starts-Offering-Super-HD-and-3D-Streams-122670}
\BIBentrySTDinterwordspacing

\bibitem{Seker.2018}
C.~Seker, M.~T. G{\"u}neser, and T.~Ozturk, ``A review of millimeter wave
  communication for 5g,'' in \emph{2018 2nd International Symposium on
  Multidisciplinary Studies and Innovative Technologies (ISMSIT), IEEE}, 2018,
  pp. 1--5.

\bibitem{Letaief.2019}
K.~B. Letaief, W.~Chen, Y.~Shi, J.~Zhang, and Y.~Zhang, ``The roadmap to 6g: Ai
  empowered wireless networks,'' \emph{IEEE Communications Magazine}, vol.~57,
  no.~8, pp. 84--90, 2019.

\bibitem{Peng.2016}
B.~Peng and T.~Kuerner, ``Three dimensional angle of arrival estimation in
  dynamic indoor terahertz channels using forward-backward algorithm,''
  \emph{IEEE Transactions on Vehicular Technology}, pp. 3791--3811, 2016.

\bibitem{Saad.2020}
W.~Saad, M.~Bennis, and M.~Chen, ``A vision of 6g wireless systems:
  Applications, trends, technologies, and open research problems,'' \emph{IEEE
  Network}, vol.~34, no.~3, pp. 134--142, 2020.

\end{thebibliography}
\bibliographystyle{IEEEtran}

\end{document}